\def\rfr#1{eq. (\ref{#1})}

\def\dert#1#2{\frac{{{d}}{#1}}{{{d}}{#2}}}              % derivate parziali e totali prima e seconda

\def\virg#1{``#1''}

\def\rf#1{Ref.~\cite{#1}}

\def\eqi{\begin{equation}}
\def\eqf{\end{equation}}
\def\eqia{\begin{eqnarray}}
\def\eqfa{\end{eqnarray}}
\def\rp#1#2{{#1\over#2}} \def\lb#1{\label{#1}}

\def\bds#1{\boldsymbol{#1}}

\documentclass[11pt]{article}%{ws-ijmpd}
\usepackage{amsmath,amsthm,amscd,amssymb}
\usepackage{latexsym}
\usepackage{graphicx,epsfig}

\begin{document}

\noindent{\bf \LARGE{Effect of Sun and Planet-Bound Dark Matter on Planet and Satellite Dynamics in the Solar System}}
\\
\\
\\
{L. Iorio$^{\ast}$\\
{\it $^{\ast}$INFN-Sezione di Pisa. Address for correspondence: Viale Unit$\grave{a}$ di Italia 68
70125 Bari (BA), Italy.  \\ e-mail: lorenzo.iorio@libero.it}

\vspace{4mm}

\begin{abstract}
We apply our recent results on orbital dynamics around a mass-varying central body to the
phenomenon of accretion of Dark Matter-assumed not self-annihilating-on the Sun and the major bodies of the solar system due to its motion throughout the Milky Way halo. We inspect its consequences  on the orbits of the planets and their satellites over timescales of the order of the age of the solar system. It turns out that a solar Dark Matter accretion rate of $\approx 10^{-12}$ yr$^{-1}$, inferred from the upper limit $\Delta M/M= 0.02-0.05$ on the Sun's Dark Matter content, assumed somehow accumulated during last 4.5 Gyr, would have displaced the planets faraway by about $10^{-2}-10^1$ au 4.5 Gyr ago. Another consequence is that the semimajor axis of the Earth's orbit, approximately equal to the Astronomical Unit, would undergo a secular increase of $0.02-0.05$ m yr$^{-1}$, in agreement with the latest observational determinations of the Astronomical Unit secular increase of $0.07\pm 0.02$ m yr$^{-1}$ and $0.05$ m yr$^{-1}$. By assuming that the Sun will continue to accrete Dark Matter in the next billions year at the same rate as in the past, the orbits of its planets will shrink by about $10^{-1}-10^1$ au ($\approx 0.2-0.5$ au for the Earth), with consequences for their fate, especially of the inner planets. \textrm{On the other hand, lunar and planetary ephemerides set upper bounds on the secular variation of the Sun's gravitational parameter $GM$ which are  one order of magnitude smaller than $\approx 10^{-12}$ yr$^{-1}$.} Dark Matter accretion on planets has, instead, less relevant consequences for their satellites. Indeed, 4.5 Gyr ago their orbits would have been just $10^{-2}-10^1$ km wider than now. Dark Matter accretion is not able to explain the observed accelerations of the orbits of some of the Galilean satellites of Jupiter, the secular decrease of the semimajor axis of the Earth's artificial satellite LAGEOS and the secular increase of the Moon's orbit eccentricity.
\end{abstract}

Keywords: Experimental studies of gravity; Dark matter;  Solar system objects; Celestial mechanics  \\
PACS: 04.80.-y, 95.35.+d, 96.30.-t, 95.10.Ce

\section{Introduction}\lb{introduzione}
Dark Matter (DM) was postulated long ago  to explain the discrepancy between the observed kinematics of some
components of astrophysical systems like clusters of galaxies \cite{Zwi} and spiral galaxies \cite{Bos,Rub}, and the predicted one on the basis of the Newtonian
dynamics and the matter directly detected from the emitted
electromagnetic radiation (visible stars and gas clouds). Zwicky \cite{Zwi2} postulated the existence of undetected, baryonic matter; today \cite{Rub83} it is
believed that the hidden mass is constituted by non-baryonic, weakly interacting
particles in order to cope with certain issues pertaining galaxy
and galaxy clusters formation \cite{Chi} and primordial nucleosynthesis \cite{Gon}. On cosmological scales, DM accounts for about $23\%$ of the mass-energy content of the Universe \cite{Hin}.

It is important to perform tests of the DM existence in other scenarios, independently of the phenomena themselves for which its existence was originally postulated. \textrm{Otherwise, it could not get rid of the unpleasant status of ad-hoc hypothesis.}

Our solar system can be regarded as one of such different laboratories. The first arguments in
support of invisible  (standard) matter in solar system were presented in 1932 by Oort \cite{Oor}  in connection with the dynamical density in the
vicinity of the Sun's Galactic orbit. Oort concluded that the total mass of nebulous or meteoric matter near the Sun is less than $3\times 10^{-24}$ g cm$^{-3}$. Nowadays, it is \textrm{phenomenologically evaluated\footnote{See later.}} from that the upper bound on local DM density in the solar system
\eqi \varrho_{\rm ss}\cong 2\times 10^{-19}\ {\rm g\ cm}^{-3} = 10^5\ \left(\rp{{\rm Gev}}{c^2}\right)\ {\rm cm^{-3}}\lb{rhoss}\eqf
is several orders of magnitude higher than
the mean Galactic value \cite{Ber} \eqi\varrho_{\rm gal}\cong 4\times 10^{-25}\ {\rm g\ cm}^{-3}= 0.3\ \left(\rp{{\rm Gev}}{c^{2}}\right)\ {\rm cm}^{-3},\lb{rhogal}\eqf so that it could be imagined that the solar system is surrounded by a local subhalo, a suggestion reinforced by recent simulations \cite{Die}. Several processes have been postulated to clump DM in the solar system \cite{Pre,Dam,Gou1,Lun,Pet1,Pet2,Pet3,Pet4}; according to \rf{Fre,Kri09}, the existence of the solar system  itself might be evidence for a local subhalo. \textrm{However, it must be noted that, according to \rf{Gou}, densities of DM as large as that of \rfr{rhoss} could not be reached because of the inverse process of capturing DM from the Milky Way halo. According to it, whenever the density of DM bound to the solar system gets big, the process of throwing WIMPs out of it will be in equilibrium with capture from the Galactic halo. This fact has been proven by using the Liouville's theorem, which says that the phase space density (after a long time, when equilibrium has been reached) of the solar system bound WIMPs is the same as in the halo \cite{Gou,Lun}. Such arguments have recently been reviewed in \cite{critica}.}
Several studies have been dedicated to placing bounds on the local distribution of DM in our solar system from orbital motions of natural major and minor bodies and artificial probes \cite{Fre,Bracci,And89,Bra,Kli,Nor,And95,Sof95,Gro,KriPit06,Ser06,Iorio06,Kri07,Fre,Adl08,Adl}\textrm{; actually, the upper bound of \rfr{rhoss} comes just from such investigations. Propagation of  electromagnetic waves has been used as well \cite{Ara}.} It has also been suggested that some proposed and approved space-based missions for fundamental physics may be used to search for local distributions of DM as well \cite{Blin,wtni,Cer}.
Apart from directly affecting orbital motions and electromagnetic waves propagation, DM may also have other effects on the solar system's bodies: for example, \textrm{it may concur to form main-sequence stars like the Sun \cite{Kar} and accrete on the major planets \cite{Xu,Kri09}, and affect their internal heats \cite{Mit,Adl09}.}
According to \rf{Xu}, during its journey along the Galaxy since its birth 4.5 Gyr ago the solar system would have encountered about  $203 M_{\odot}$ of DM. In \rf{Kar} it is stated that the upper limit to the amount of gravitating DM in the Sun would be $2-5\%$ of the total solar mass; by assuming that it somehow accumulated during the Sun's lifetime, it is possible to infer a mass increase rate of \eqi\rp{\dot M}{M}=4.4-11\times 10^{-12}\ {\rm yr}^{-1}.\lb{rate1}\eqf
\textrm{In fact, in \rf{Kar} it is argued that the DM content of the Sun cannot be due to a continuous accretion throughout its lifetime; however, the value of the amount of Galactic DM encountered by the Sun in the latest 4.5 Gyr in \rf{Kar} radically differs from that in \rf{Xu}.}

\textrm{Accretion of DM, in the form of Mirror Matter \cite{klobook}, by the Sun during its lifetime is envisaged in \rf{Blin}.
Indeed, according to eq. (7) of \rf{Blin}, reproduced below, a gravitating body of mass $M$ traveling at speed $v$ through a volumetric distribution of DM with density $\varrho$ will accrete an amount $\Delta M$ over a time span $t$
\eqi \rp{\Delta M}{M}\simeq 10^{-5}\left(\rp{M}{M_{\odot}}\right)\left(\rp{v}{10\ {\rm km\ s}^{-1}}\right)^{-3}\left(\rp{\varrho}{10^{-24}\ {\rm g\ cm}^{-3}}\right)\left(\rp{t}{10^{10}\ {\rm yr}}\right).\lb{BLIN}\eqf Thus, if we assume $v=30$ km s$^{-1}$ and the upper limit of the local DM density in the solar system given by \rfr{rhoss}, the Sun will increase its mass at a rate
\eqi \rp{\dot M}{M}\simeq 7 \times 10^{-12}\ {\rm yr}^{-1},\lb{ratazzo}\eqf close to \rfr{rate1} from \rf{Kar}.
On the other hand, the mean Galactic value of \rfr{rhogal} applied to \rfr{BLIN} yields a mass accretion rate of
\eqi \rp{\dot M}{M}\simeq 1.5 \times 10^{-17}\ {\rm yr}^{-1}.\lb{ratezzo}\eqf
For several possible observable manifestations of Mirror Matter at various astronomical scales, see \rf{Khlop}.}

Concerning the planets, according to Table II and Table III of \rf{Xu}, Jupiter would have captured the largest amount of DM during the latest 4.5 Gyr followed by Saturn, Uranus  and Neptune. See Table \ref{dmdt} for the related relative mass variation rates $\dot M/M$ of the eight planets.

\begin{table}[!ht]
\caption{DM accretion rates $\dot M/M$ on major bodies of the solar system according to Table II and Table III of \protect\rf{Xu}. }\label{dmdt}
%\centering
%
%\bigskip
\begin{tabular}{ll}
\hline\noalign{\smallskip}
Planet & $\dot M/M $ (yr$^{-1}$) \\
\noalign{\smallskip}\hline\noalign{\smallskip}
Mercury & $ 2.7\times 10^{-17}$\\
Venus & $1.6\times 10^{-17}$\\
Earth & $1.4\times 10^{-17}$\\
Mars & $4.1\times 10^{-17}$\\
Jupiter & $0.6\times 10^{-17}$\\
Saturn & $1.1\times 10^{-17}$\\
Uranus & $2.9\times 10^{-17}$\\
Neptune & $3.5\times 10^{-17}$\\
\noalign{\smallskip}\hline\noalign{\smallskip}
\end{tabular}
\end{table}

The estimates in \rf{Kri09} are different but, as we will show, such discrepancies will not substantially affect the main result of our analysis.

We will investigate if the phenomenon of DM capture and accretion
by the Sun and the planets has somewhat influenced the solar system dynamics throughout its 4.5 Gyr lifetime.
We will address this problem in a specific way, i.e. we will look at the possible modifications
of the orbital trajectories of the satellites of the major bodies which experienced the DM accretion, assumed not self-annihilating \cite{Adl09}.

The paper is organized as follows. In Section \ref{due} we will briefly review some basics of orbital motion around a mass-varying body.
In Section \ref{tre} we will apply our results to concrete astronomical scenarios in the solar system.
Section \ref{quattro} is devoted to the Conclusions.
\section{Orbital dynamics around a mass-varying body}\lb{due}
The acceleration experienced by a test particle orbiting a mass-varying central body with $\mu(t)\doteq GM(t)$ can be approximated
with
\eqi \bds A = -\rp{\mu(t)}{r^2}\bds{\hat{r}}\approx-\rp{\mu}{r^2}\left[1+\left(\rp{\dot\mu}{\mu}\right)(t-t_0)\right]\bds{\hat{r}},\lb{acce}\eqf
where $t_0$ is a given epoch and
$\mu\doteq \left.\mu\right|_{t=t_0},\ \dot\mu\doteq \left.\dot\mu\right|_{t=t_0}.$ In our case $t_0$ coincides with the present time, and $t< t_0$; thus, $t-t_0<0$. Since in our case $\dot \mu$ is due to DM accretion, we will assume that $\dot\mu=$cost. throughout the solar system's lifetime, i.e. we will not deal with $\ddot\mu$; the validity of such an approximation will be justified a-posteriori in Section \ref{tre}. Moreover, the condition of validity of \rfr{acce}, i.e. $(\dot\mu/\mu)(t-t_0)\ll 1$, is  satisfied for all the major bodies of the solar system over $\Delta t=-4.5$ Gyr, given the DM accretion rates quoted in Section \ref{introduzione}.

In \rf{Iorio10} the secular, i.e. averaged over one orbital revolution, orbital effects induced by \rfr{acce} on the usual Keplerian orbital elements\footnote{The type I, according to \rf{Khol}} of a test particle have been analytically worked out by means of the Gauss equations for the variations of elements. They are
\begin{eqnarray}
  \left\langle \dot a \right\rangle &=& \rp{2ea}{(1-e)}\left(\rp{\dot\mu}{\mu}\right) \lb{sma}\\
  \left\langle \dot e \right\rangle &=& (1+e)\left(\rp{\dot\mu}{\mu}\right) \lb{ecc}\\
   \left\langle\dot I \right\rangle &=& 0 \\
   \left\langle\dot\Omega \right\rangle &=& 0 \\
   \left\langle\dot\omega \right\rangle &=& 0 \\
  \left\langle\dot{\mathcal{M}} \right\rangle &=& n + 2\pi\left(\rp{\dot\mu}{\mu}\right),
\end{eqnarray}
where $a,e,I,\Omega,\omega,{\mathcal{M}}$ are the values at the epoch of the semimajor axis, the eccentricity, the inclination, the longitude of the ascending node, the argument of pericenter and the mean anomaly, respectively; $n\doteq \sqrt{\mu/a^3}$ is the Keplerian mean motion. Such results are not in contrast with the  orbital evolution effectively followed by the test particle, as shown in detail in \rf{Iorio10}. Indeed, for, e.g., $\dot\mu/\mu<0$, i.e. for a mass decrease, the trajectory, which is not an ellipse, would expand in such a way that the osculating semimajor axis and eccentricity  of the Keplerian ellipses approximating it at the pericentre, which remains fixed, get reduced revolution after revolution, in accordance to \rfr{sma} and \rfr{ecc}; see Figure 1 of \rf{Iorio10} and the related discussion. Moreover, also the osculating Keplerian period $P_{\rm b}\doteq 2\pi/n$, i.e. the time required to fully describe an osculating Keplerian ellipse, gets shorter, according to
\eqi \left\langle \dert{P_{\rm b}} t \right\rangle = \rp{3}{2}P_{\rm b}\rp{\left\langle\dot a \right\rangle}{a},\lb{perio}\eqf while the time of two consecutive pericentre crossings along the true path grows.

The variations of the radial $\rho$ and transverse $\tau$ components of the radius vector, taken at two consecutive pericentre passages, are \cite{Iorio10}
\begin{eqnarray}
\Delta \rho &=& -\rp{2\pi}{n}a(1-e)\left(\rp{\dot\mu}{\mu}\right),\lb{ra}\\
\Delta\tau &=& \rp{4\pi^2}{n}a\left(\rp{\dot\mu}{\mu}\right)\sqrt{\rp{1+e}{1-e}},\lb{tra}
\end{eqnarray}
so that the overall pericenter-to-pericenter orbit variation is
\eqi\Delta r= \Delta \rho\sqrt{1+4\pi^2\rp{(1+e)}{(1-e)^3}}.\lb{tota}\eqf
\section{Consequences of DM accretion on the orbits of the satellites of the major planets}\lb{tre}
Here we will fruitfully apply the results of Section \ref{due} to the problem of determining the impact of DM accretion on the Sun and the major planets  on the orbital evolution of their planetary and satellite systems throughout the past 4.5 Gyr.
\subsection{The Sun and the planets}
Concerning the Sun, by assuming for it $\dot M/M=4.4-11\times 10^{-12}$ yr$^{-1}$ \cite{Kar}, it turns out that 4.5 Gyr ago the orbits of the planets were larger than now by the amounts listed in Table \ref{prima}.
\begin{table}[!ht]
\caption{Increment $\Delta r$ of the planetary orbits for $\dot M/M=4.4-11\times 10^{-12}$ yr$^{-1}$ \protect\cite{Kar} and $\Delta t=-4.5$ Gyr according to \rfr{tota}. }\label{prima}
%\centering
%
%\bigskip
\begin{tabular}{ll}
\hline\noalign{\smallskip}
Planet & $\Delta r$ (au) \\
\noalign{\smallskip}\hline\noalign{\smallskip}
Mercury & $0.06-0.15$\\
Venus & $0.09-0.23$\\
Earth & $0.13-0.32$\\
Mars & $0.21-0.52$\\
Jupiter & $0.68-1.71$\\
Saturn & $1.26-3.17$\\
Uranus & $2.53-6.31$\\
Neptune & $3.82-9.55$\\
\noalign{\smallskip}\hline\noalign{\smallskip}
\end{tabular}
\end{table}
It can be noted that, while the distances of the inner planets at their formation are not substantially altered, the giant gaseous planets are shifted by about $1-10$ au with respect to their present-day locations. If the rate of DM accretion on the Sun will be the same for the remaining lifetime of the solar system, the consequences for its planets would be relevant because of the continuous shrinking of their orbits. Indeed, when the Sun will have reached the tip of the Red Giant Branch (RGB) in the Hertzsprung-Russel (HR) diagram in the next 7.58 Gyr \cite{RGB}, the orbits of the planets will be reduced by the amounts listed in Table \ref{shrink}.
\begin{table}[!ht]
\caption{Decrement $\Delta r$ of the planetary orbits for $\dot M/M=4.4-11\times 10^{-12}$ yr$^{-1}$ \protect\cite{Kar} and $\Delta t=+7.58$ Gyr according to \rfr{tota}. }\label{shrink}
%\centering
%
%\bigskip
\begin{tabular}{ll}
\hline\noalign{\smallskip}
Planet & $\Delta r$ (au) \\
\noalign{\smallskip}\hline\noalign{\smallskip}
Mercury & $-(0.10-0.25)$\\
Venus & $-(0.15-0.38)$\\
Earth & $-(0.21-0.54)$\\
Mars & $-(0.35-0.88)$\\
Jupiter & $-(1.15-2.89)$\\
Saturn & $-(2.13-5.33)$\\
Uranus & $-(4.25-10.65)$\\
Neptune & $-(6.43-16.08)$\\
\noalign{\smallskip}\hline\noalign{\smallskip}
\end{tabular}
\end{table}
In particular, the heliocentric distance of the Earth may be reduced by $\approx 0.2-0.5$ au. Figure 2 of \rf{RGB} shows that at the beginning of RGB the solar photosphere will reach about $0.5-0.6$ au; after entering the RG phase things will dramatically change because in only $\approx 1$ Myr the Sun will reach the tip of the RGB phase expanding up to 1.20 au, while the Earth's distance will be just $0.8-0.5$ au or less. However, it must be pointed out that the results of \rf{RGB} do not take into account possible DM accretion which, in principle, may have consequences on the lifetime cycle of main sequence stars  depending on the type of DM involved \cite{Mstar}.
Another interesting consequence of the Sun's DM accretion is that the semimajor axis $a_{\oplus}$ of the Earth's orbit, which can be approximately taken equal to the Astronomical Unit amounting to\footnote{See http://ssd.jpl.nasa.gov/txt/p$\_$elem$\_$t2.txt on the WEB.} $a_{\oplus} = 1.00000018$ au, undergoes a secular increment of just $0.02-0.05$ m yr$^{-1}$ according to \rfr{sma} and $\dot M/M=4.4-11\times 10^{-12}$ yr$^{-1}$.
An anomalous increase of the Astronomical Unit of $0.15\pm 0.04$ m yr$^{-1}$  was reported for the first time in \rf{Kra}.
Later results are $0.07\pm 0.02$ m yr$^{-1}$ \cite{StaAU} and $0.05$ m yr$^{-1}$ \cite{PitAU}.
\textrm{Anyway, we point out that the figures presented here would be radically smaller if \rfr{rhogal} for the Galactic mean density was used in \rfr{BLIN} of \rf{Blin}.
An independent check of the assumption on which the previously results are based, i.e. the Sun's mass accretion rate of \rfr{rate1}, comes from the upper bounds on the secular variation of the Newtonian constant of gravitation $G$ which have been recently obtained by processing long data sets of planetary observations with the latest ephemerides. Actually, they have been derived from the variation of the gravitational parameter $\mu\doteq GM$ of the Sun, but
\eqi \rp{\dot\mu}{\mu}=\rp{\dot G}{G}+\rp{\dot M}{M},\eqf so that they are useful for our purposes as well. W.M. Folkner, by analyzing the Mars spacecraft data set with the JPL DE421 ephemerides recently obtained \cite{DE421}
\eqi \left|\rp{\dot\mu}{\mu}\right|\leq 2\times 10^{-13}\ {\rm yr}^{-1}.\eqf Lunar Laser Ranging yields \cite{LLR}
\eqi \rp{\dot\mu}{\mu}=(4\pm 9) \times 10^{-13}\ {\rm yr}^{-1},\eqf with an upper bound of $1.3\times 10^{-12}$ yr$^{-1}$, and \cite{Bisku}
\eqi \rp{\dot\mu}{\mu}=(2\pm 7) \times 10^{-13}\ {\rm yr}^{-1},\eqf yielding an upper bound of $9\times 10^{-13}$ yr$^{-1}$.
Let us mention that E.V. Pitjeva, with the EPM ephemerides, recently obtained a statistically significative non-zero result \cite{Pit010}
\eqi \rp{\dot\mu}{\mu}=(-5.9\pm 4.4) \times 10^{-14}\ {\rm yr}^{-1};\eqf the upper bound is $-1.5\times 10^{-14}$ yr$^{-1}$.
Such results contradict the possibility that a steady mass accretion as large as that by \rfr{rate1}-\rfr{ratazzo} may occur for the Sun; on the contrary, the smaller value by \rfr{ratezzo} due to the mean Galactic DM density would be allowed.
}
\subsection{The planets and their satellites}
Let us, now, move to the planetary systems of satellites. In the case of the Earth and Moon, \rfr{tota} and $\dot M/M=1.4\times 10^{-17}$ yr$^{-1}$ from \rf{Xu} yields an expansion of the lunar orbit of just 0.16 km. The recently observed secular increase  of the eccentricity of the geocentric Moon's orbit amounts to $\left\langle\dot e\right\rangle = (0.9\pm 0.3)\times 10^{-11}$ yr$^{-1}$ \cite{moon}. It is several orders of magnitude larger that the secular increase predicted by \rfr{ecc}. Concerning the observationally determined secular decrement $\dot a=-0.4$ m yr$^{-1}$ of the semimajor axis of the orbit of the geodetic satellite LAGEOS \cite{Ruby}, DM accretion is unable to explain it because \rfr{sma} yields an increase of the order of $10^{-12}$ m yr$^{-1}$.

In Table \ref{Giove}-Table \ref{Nettuno} we quote the expansion of the orbits of some satellites of Jupiter, Saturn, Uranus and Neptune 4.5 Gyr ago induced by
the DM accretion according to \rf{Xu}.
\begin{table}[!ht]
\caption{Jupiter: increment $\Delta r$ of the Galilean satellites' orbits and of S/2003J2 for $\dot M/M=0.6\times 10^{-17}$ yr$^{-1}$ \protect\cite{Xu} and $\Delta t=-4.5$ Gyr according to \rfr{tota}. }\label{Giove}
%\centering
%
%\bigskip
\begin{tabular}{ll}
\hline\noalign{\smallskip}
Satellite & $\Delta r$ (km) \\
\noalign{\smallskip}\hline\noalign{\smallskip}
Io & 0.07\\
Europa & 0.11\\
Ganymede & 0.17\\
Callisto & 0.31\\
S/2003J2 & 7.21\\
\noalign{\smallskip}\hline\noalign{\smallskip}
\end{tabular}
\end{table}
\begin{table}[!ht]
\caption{Saturn: increment $\Delta r$ of some satellites'orbits for $\dot M/M=1.1\times 10^{-17}$ yr$^{-1}$ \protect\cite{Xu} and $\Delta t=-4.5$ Gyr according to \rfr{tota}. }\label{Saturno}
%\centering
%
%\bigskip
\begin{tabular}{ll}
\hline\noalign{\smallskip}
Satellite & $\Delta r$ (km) \\
\noalign{\smallskip}\hline\noalign{\smallskip}
Mimas & 0.06\\
Titan & 0.39\\
Iapetus & 1.15\\
Phoebe & 4.78\\
Fornjot & 9.73\\
\noalign{\smallskip}\hline\noalign{\smallskip}
\end{tabular}
\end{table}
\begin{table}[!ht]
\caption{Uranus: increment $\Delta r$ of some satellites'orbits for $\dot M/M=2.9\times 10^{-17}$ yr$^{-1}$ \protect\cite{Xu} and $\Delta t=-4.5$ Gyr according to \rfr{tota}. }\label{Urano}
%\centering
%
%\bigskip
\begin{tabular}{ll}
\hline\noalign{\smallskip}
Satellite & $\Delta r$ (km) \\
\noalign{\smallskip}\hline\noalign{\smallskip}
Ariel & 0.15\\
Oberon & 0.48\\
Margaret & 26.4\\
Setebos & 28.18\\
Ferdinand & 25.62 \\
\noalign{\smallskip}\hline\noalign{\smallskip}
\end{tabular}
\end{table}
\begin{table}[!ht]
\caption{Neptune: increment $\Delta r$ of some satellites'orbits for $\dot M/M=3.5\times 10^{-17}$ yr$^{-1}$ \protect\cite{Xu} and $\Delta t=-4.5$ Gyr according to \rfr{tota}. }\label{Nettuno}
%\centering
%
%\bigskip
\begin{tabular}{ll}
\hline\noalign{\smallskip}
Satellite & $\Delta r$ (km) \\
\noalign{\smallskip}\hline\noalign{\smallskip}
Naiad & 0.05\\
Galatea & 0.06\\
Triton & 0.35\\
Nereid & 14.46\\
\noalign{\smallskip}\hline\noalign{\smallskip}
\end{tabular}
\end{table}
It can be shown that the effect of such form of $\dot M/M$ is very tiny, amounting approximately to $10^{-2}-10^2$ km. This shows that the discrepancies between the evaluations of \rf{Xu} and \rf{Kri09} are negligible in our case. Moreover, a-posteriori we can well justify the assumption of neglecting $\ddot\mu$.
Let us note that DM accretion on Jupiter cannot be the cause of the observationally determined acceleration of the orbits of Io, Europa and Ganymede \cite{jupio}, because the orbital period of Io is shortening, while those of Europa and Ganymede are lengthening; \rfr{perio}, instead, predicts an uniform increase for $P_{\rm b}$.
We can conclude that DM accretion on the major planets of the solar system had negligible impact in sculpting the orbital patterns of their satellite systems.
\section{Conclusions}\lb{quattro}
In this paper we have investigated the impact that the phenomenon of DM capture by the Sun and the major bodies of the solar system may have on the orbital dynamics of the planets and their satellites on timescales comparable with the age of the solar system.

Concerning the planets, a solar mass increase by DM-assumed not-self-annihilating-as large as $\approx 10^{-12}$ yr$^{-1}$ would displace their orbits outward by about $10^{-2}-10^1$ au at the epoch of the solar system's formation 4.5 Gyr ago; for the Earth the increase of its heliocentric distance would be in the range $0.1-0.3$ au, while the largest shifts are for the giant gaseous planets. \textrm{The solar DM accretion rate is assumed by postulating that the maximum amount of DM in the Sun, evaluated as $2-5\%$ of the total solar mass, comes from a continuous accretion process. Anyway, we note that the real occurrence of such a process is controversial; moreover, also in this case, it would be orders of magnitude smaller if the local DM density was as large as the mean Galactic one.}
\textrm{On one hand,} one of the consequences of the assumed DM-induced solar accretion is that the osculating semimajor axis of the Earth's shrinking orbit would experience a secular increase as large as the latest observational determinations of the Astronomical Unit rate of $\approx 0.07-0.05$ m yr$^{-1}$. In fact, no real contradiction exists because it has been shown that the osculating semimajor axis of a gradually shrinking trajectory around a mass-increasing central body gets, in fact, larger.
By assuming that the solar DM accretion will take place in the next billions year at the same rate as in the past, the orbits of the planets will shrink by $\approx 10^{-1}-10^1$ au ($0.2-0.5$ au for the Earth), with consequences for their fate, especially for the inner planets.
\textrm{On the other hand, long observational data records processed with the latest lunar and planetary ephemerides set upper bounds on possible secular variations of the Sun's gravitational parameter $GM$ which are more than one order of magnitude smaller than $10^{-12}$ yr$^{-1}$.
}

The DM accretion on the planets has, instead, no effects on the dynamics of their satellites whose planetocentric distances are altered by just $10^{-2}-10^1$ km over 4.5 Gyr. Concerning the Earth, both the observationally determined secular decrease of the semimajor axis of the LAGEOS satellite of $0.4$ m yr$^{-1}$ and the secular increase of the eccentricity of $(0.9\pm 0.3)\times 10^{-11}$ yr$^{-1}$ of the lunar orbit cannot be explained by the DM accretion on the Earth. The same occurs for the observed accelerations of the orbits of Io, Europa and Ganymede.

\section*{Acknowledgments}
I thank M.Yu. Khlopov  for useful references and comments.

\end{document}